# Multiphase strontium molybdate thin films for plasmonic local heating applications

Matthew P. Wells[1], Bin Zou[1], Andrei P. Mihai[1], Ryan Bower[1], Anna Regoutz[1], Sarah Fearn[1], Stefan A. Maier[2,3], Neil McN. Alford[1], Peter K. Petrov[1]*

[1] *Imperial College London, Department of Materials, Prince Consort Road, London SW7 2BP, UK*
[2] *Imperial College London, Department of Physics, Prince Consort Road, London SW7 2AZ, UK*
[3] *Ludwig-Maximilians-Universität München, Faculty of Physics, 80799 München, Germany*
*\*p.petrov@imperial.ac.uk*

**Abstract:** In the search for alternative plasmonic materials $SrMoO_3$ has recently been identified as possessing a number of desirable optical properties. Owing to the requirement for many plasmonic devices to operate at elevated temperatures however, it is essential to characterize the degradation of these properties upon heating. Here, $SrMoO_3$ thin films are annealed in air at temperatures ranging from 75 - 500° C. Characterizations by AFM, XRD, and spectroscopic ellipsometry after each anneal identify a loss of metallic behaviour after annealing at 500° C, together with the underlying mechanism. Moreover, it is shown that by annealing the films in nitrogen following deposition, an additional crystalline phase of $SrMoO_4$ is induced at the film surface, which suppresses oxidation at elevated temperatures.

## 1. Introduction

In the pursuit of improved technologies for energy harvesting [1-4], communication systems [5, 6], and sub- wavelength imaging [7, 8], among many others [9-12] the field of plasmonics is one which holds a great deal of promise. This is a result of the means by which the energy of incident light may be efficiently coupled to the free electrons at a metal/dielectric interface. However, there currently exist a number of barriers preventing the widespread realisation of plasmonic devices. For example, the noble metals Au and Ag, which have formed the basis of a great deal of plasmonics research, owing to their high electrical conductivity, are incompatible with currently ubiquitous CMOS technologies [13]. Moreover, although Au exhibits superior chemical and thermal stability compared to Ag, the optical properties of the material are subject to significant variation on exposure to temperatures in excess of approximately 300° C, thus preventing its use in high temperature applications such as heat - assisted magnetic recording (HAMR) and solar thermophotovoltaics (STPV) [14, 15].

The encouraging plasmonic properties of the perovskite material strontium molybdate ($SrMoO_3$, SMO), namely its strain-dependent tunable optical properties and relatively low optical losses, have been recently reported [16]. However, it was noted that some concerns exist regarding the material's propensity to oxidise and hence its applicability in real-world applications, in particular those operating at elevated temperatures. In the present work, the use of an in-situ annealing process is proposed to introduce an additional crystalline phase ($SrMoO_4$) to the SMO thin film. Samples were produced by pulsed laser deposition (PLD) onto (100) oriented strontium titanate ($SrTiO_3$, STO) substrates at 650° C under vacuum conditions. Following film growth, the vacuum chamber was filled with $N_2$ gas (6N purity) to a pressure of 500 Torr and annealed, also at 650° C, for varying time periods. Samples were subsequently characterised by means of X-ray diffraction (XRD), X-ray photoelectron spectroscopy (XPS), secondary-ion mass spectrometry (SIMS), atomic force microscopy (AFM), spectroscopic ellipsometry, and DC resistivity measurements, prior to ex-situ annealing in air to study the behaviour of the films at elevated temperatures.

## 2. Experimental

The SrMoO$_3$ PLD target material was prepared from SrMoO$_4$ powder (99.9% purity) supplied by Alfa Aesar. The powder was placed in propan-2-ol and ball milled at 300 rpm for 20 hrs before evaporating the propan-2-ol by placing the powder in an oven overnight at 60° C. The powder was then reduced in a furnace under 100 mL min$^{-1}$ gas flow of 5% H$_2$ / 95% N$_2$ at 1400° C for 10 hrs. The powder was then pressed into a target with a density of approximately 4 g cm$^{-3}$ before sintering under the same gas flow conditions at 1500° C for 12 hours.

The SrMoO$_3$ target material was rotated throughout each pulsed laser deposition process and held 60 mm from the SrTiO$_3$ substrate. A KrF excimer laser (240 nm) was used for the deposition of all samples with a repetition rate of 8 Hz and a 10 s relaxation period after every 20 pulses. A laser fluency of 1.2 J cm$^{-2}$ was used. Vacuum conditions, approximately 1×10$^{-7}$ Torr, were used for the deposition of all samples and the substrate temperature was 650° C. The samples were cooled to room temperature after each deposition process at a rate of 10° C min$^{-1}$, either in vacuum or following annealing in 500 Torr N$_2$ (6N purity, supplied by BOC), prior to removal from the vacuum chamber. Single side polished 5x5 mm (100) oriented STO substrates with a thickness of 0.5 mm were used.

An IONTOF ToF-SIMS 5 instrument was used for SIMS depth profiling of the samples. An area of 100x100 μm$^2$ was analysed using a 25 keV Bi+ LMIG in high current bunch mode with a beam current of approximately 1 pA. Only negative secondary ions were collected. For depth profiling, a 1 keV Cs+ ion beam with a current of 75 nA was used, giving a sputter crater area of 300x300 μm$^2$.

The surfaces of the SMO films before and after annealing were characterised using X-ray photoelectron spectroscopy (XPS). The spectra were recorded on a Thermo Scientific K-Alpha+ spectrometer operating at a base pressure of 2x10$^{-9}$ mbar. This system incorporates a monochromated, microfocused Al Kα X-ray source (hν = 1486.6 eV) and a 180° double focusing hemispherical analyser with a 2D detector. The X-ray source was operated a 6 mA emission current and 12 kV anode bias. Data were collected at pass energies of 200 eV for survey, and 20 eV for core level spectra using an X-ray spot size of 400 µm. Samples were mounted using carbon loaded conductive tape and, in addition, a flood gun was used to minimise sample charging. All spectra were aligned using the C 1s contribution of adventitious carbon at 285.0 eV. All data were analysed using the Avantage software package.

Other sample characterisations were conducted ex-situ at room temperature. Film thickness measurements were taken using a Dektak 150 surface profiler; X-ray diffraction measurements with a Bruker D2 PHASER system (Cu Kα wavelength of 1.54 Å); and spectroscopic ellipsometry measurements using a J. A. Woollam Co. HS-190 ellipsometer at a 65-80° angle of incidence. The properties of the SMO films were directly fitted to experimental data using the Marquardt minimisation algorithm while the optical constants of the STO substrates were considered known.

## 3. Results and discussion

Following film growth, the samples were first characterised by XRD. Figure 1 shows the results from two samples, one of which was cooled to room temperature under vacuum conditions after deposition, and another which was annealed in 500 Torr N$_2$ for 2 hours before cooling to room temperature.

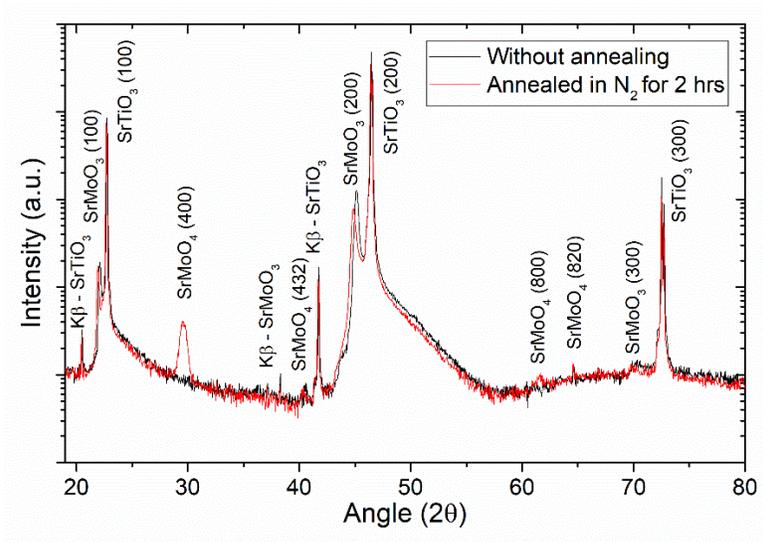

Fig. 1. XRD patterns for strontium molybdate samples produced with and without annealing in nitrogen.

The above results show that, while the non-annealed sample exhibits a single crystalline phase of SrMoO$_3$, the N$_2$ annealing process gives rise to additional SrMoO$_4$ peaks, thus indicating that the annealing process enables the crystallisation of a higher oxygen state. It may be observed that the SrMoO$_4$ peaks are broader than those of the SrMoO$_3$, indicating a lower degree of crystallinity for the additional phase. This process is further characterised by SIMS measurements, the results of which are shown in Figure 2. Here, one may observe that the annealing process acts to promote the molybdenum species of higher oxidation states towards the surface of the film, whereupon the externally high N$_2$ pressure encourages the formation of a crystalline layer, while preventing excess oxygen from diffusing away from the sample. Thus, resulting in the broad SrMoO$_4$ (400) peak observed in the XRD spectrum.

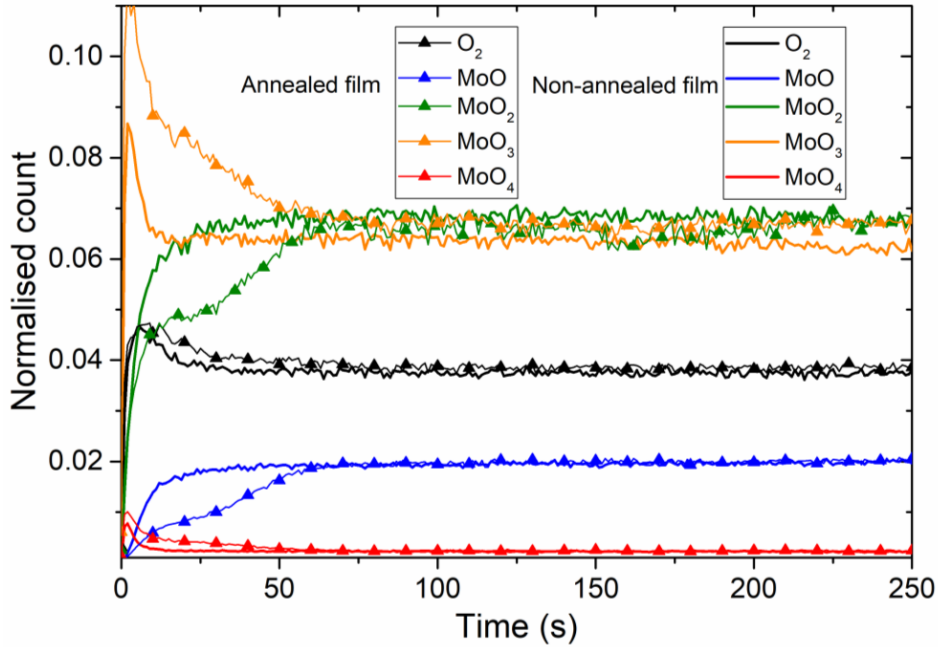

Fig. 2. SIMS profiles for SMO samples with and without annealing in $N_2$ show that the annealed film has a higher concentration of $MoO_3$ and $MoO_4$ at the sample surface than the non-annealed film. As sputter time increases beyond approximately 75 s, it can be seen that the bulk composition of the two samples is unchanged by the annealing process.

Similarly, characterisation of the surface of the SMO thin films using XPS confirms the formation of $SrMoO_4$ after annealing. Whilst the non-annealed film shows a wide range of Mo oxidation states on the surface (see Figure 3), exemplary data from a sample annealed for 1 hr show the Mo 3d core level spectrum is dominated by a $Mo^{6+}$ peak at 232.8 eV with only minor contributions from lower binding energy (BE) species. From simple peak fit analysis the ratios of $Mo^{6+}$ in $SrMoO_4$ to other Mo species can be derived, giving values of 94 : 6 (±0.5) for the annealed film, and 38 : 62 (±5) for the non-annealed film, respectively. The O 1s core level further supports this observation, showing a single peak at 530.5 eV for the annealed sample, while the non-annealed sample shows a more complex line shape with at least three contributions from varying oxygen environments. The Sr 3d core level again shows a single $SrMoO_4$ environment for the annealed film, while the non- annealed film shows an additional shoulder towards higher BE (marked with an asterisk) which is due to formation of a surface carbonate common on Sr oxide surfaces.

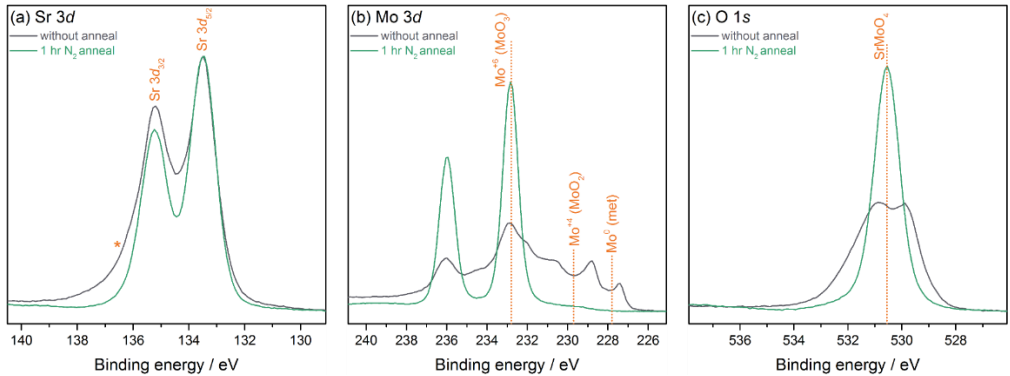

Figure 3. XPS core level spectra of strontium molybdate thin films with and without annealing in $N_2$, including (a) Sr 3d, (b) Mo 3d, and (c) O1s. All spectra are normalised to the respective Sr intensity.

Along with significant chemical changes, the annealing process was also found to result in structural changes to the films. This is indicated by Figure 4, in which AFM images of a 1x1 μm² area are shown for samples without annealing (Figure 3a) and after annealing in $N_2$ for one hour (Figure 3b).

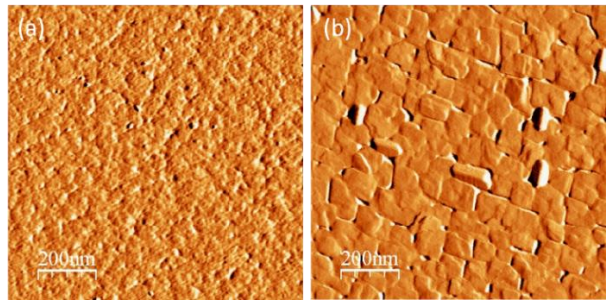

Fig. 4. AFM images for samples produced (a) without annealing and (b) with 1 hr annealing in nitrogen.

The effects of these chemical and structural changes on the optical characteristics of SMO are described in Figure 5, in which the real and imaginary parts of the dielectric permittivity are shown for samples annealed for time periods ranging from 0 to 5 hours.

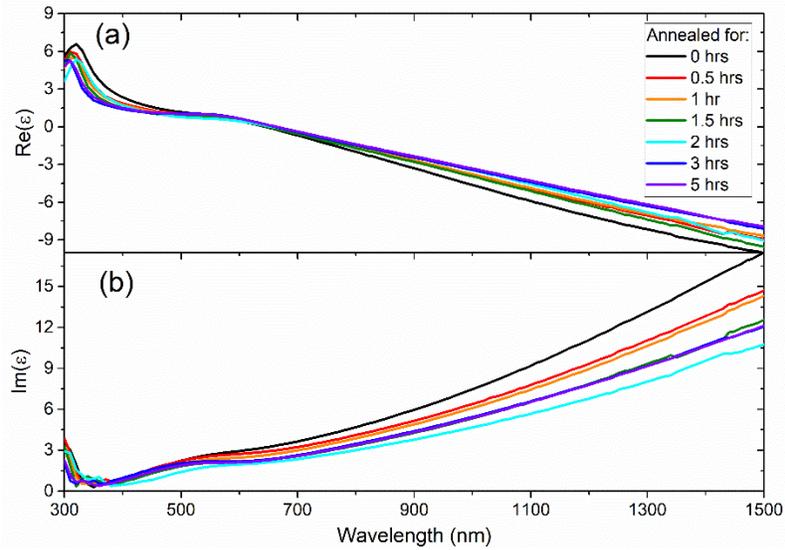

Fig. 5. Real (a) and imaginary (b) parts of the dielectric permittivity for samples annealed in nitrogen for 0-5 hrs.

SMO has already been noted for its relatively low optical loss characteristics [16]; from Figure 5 it can be seen that losses tend to decrease further with increasing annealing time. However, a minimum value is observed with an annealing time of two hours, with relatively little change observed for longer annealing times. Furthermore, the in-situ annealing process is found to have relatively little impact on the real part of the dielectric permittivity of samples. Quantifiably, comparing the non-annealed sample and the sample annealed for two hours, it can be shown that the in-situ annealing process is responsible for a 32.8% drop in Im(ε) (indicative of the optical losses) at the screened plasma frequency, $\omega_p$; however $\omega_p$ itself is reduced by only 2.9%. The reduction in the imaginary part of the dielectric permittivity may be correlated with an increase in the intensity of the $SrMoO_4$ (400) peak, taken relative to that of the SMO (200) peak as shown in Figure 6 below, in which the relationship between the $SrMoO_4$ (400) peak intensity and annealing time is also described.

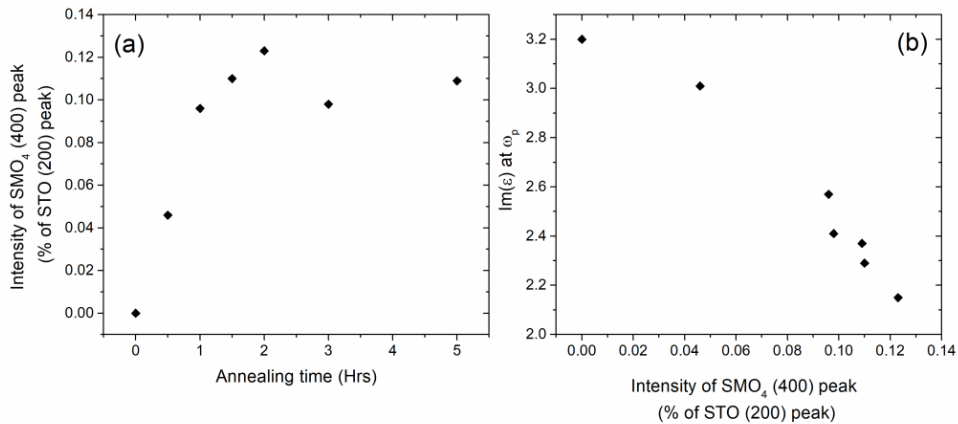

Fig. 6. (a) Variations in $SrMoO_4$ peak intensity as a function of annealing time in $N_2$ and (b) effect of increasing $SrMoO_4$ peak intensity on optical losses.

To determine any impact that the in-situ annealing may have on the thermal stability of SMO, the degradation mechanisms of a non-annealed sample were first characterised. The sample, with a thickness of approximately 75 nm, was annealed, ex situ, in air for one hour at temperatures raised incrementally to a maximum of 500° C, with analysis conducted by XRD, AFM, and spectroscopic ellipsometry after each anneal. Figure 7 shows the ellipsometry measurements taken after annealing at each temperature.

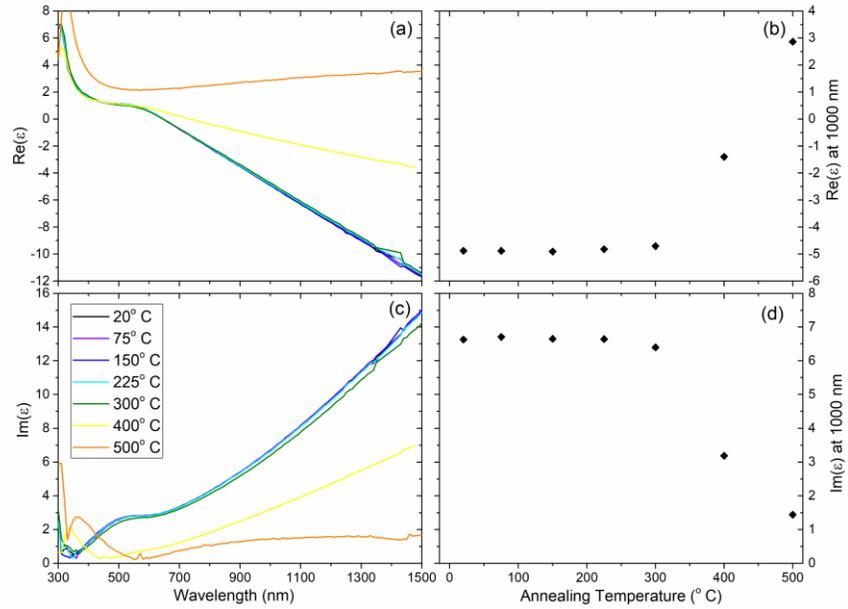

Fig. 7. Real (a) and imaginary (c) parts of the dielectric permittivity for sample annealed in air at temperatures up to 500° C. (b) and (d) show respectively the values of the real and imaginary parts of the permittivity measured at 1000 nm after each annealing process.

The above figure shows clearly that, while the optical properties of SMO are unchanged up to approximately 300° C, above this temperature the dielectric permittivity tends towards dielectric behaviour as Re($\varepsilon$) becomes positive and Im($\varepsilon$) tends towards zero. Furthermore, Figure 8 describes the significant changes observed in both the XRD spectra (a) and AFM measurements (b). It can be seen that, upon annealing at temperatures greater than 300° C, there is a loss of crystallinity (indicated by the decrease in intensity of the SMO (200) peak relative to that of the STO (200)) together with an increase in surface roughness, leading to the observed loss of metallic behaviour.

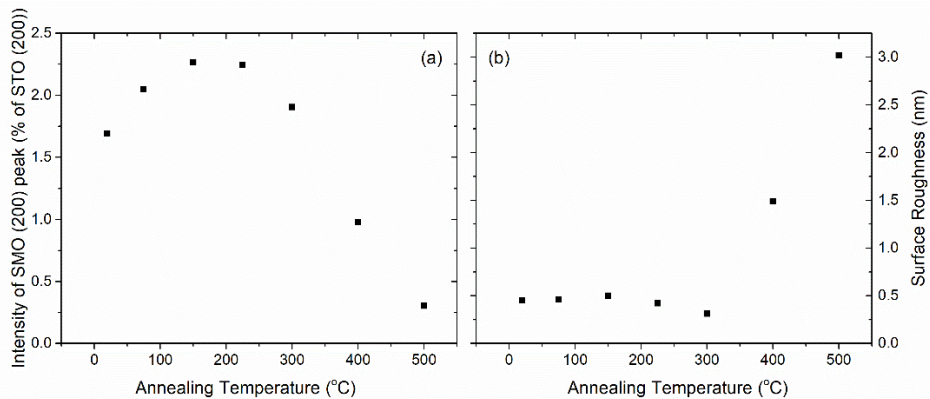

Fig. 8. (a) Variation of the intensity of SMO (200) peak as a function of annealing temperature and (b) variation in surface roughness as a function of annealing temperature.

The above results serve to highlight the instability of SrMoO$_3$, particularly when compared with other plasmonic materials such as Au, which has been shown to retain its metallic properties in air up to 500° C, albeit with some variation in plasma frequency and loss characteristics [14].

Recalling the previous results of Figures 1 - 4, an additional sample, also approximately 75 nm in thickness, which had been annealed in N$_2$, in-situ, for one hour was subject to the same ex-situ annealing process in air, to establish whether the additional crystalline phase and differing microstructure might serve to suppress the degradation mechanisms described above.

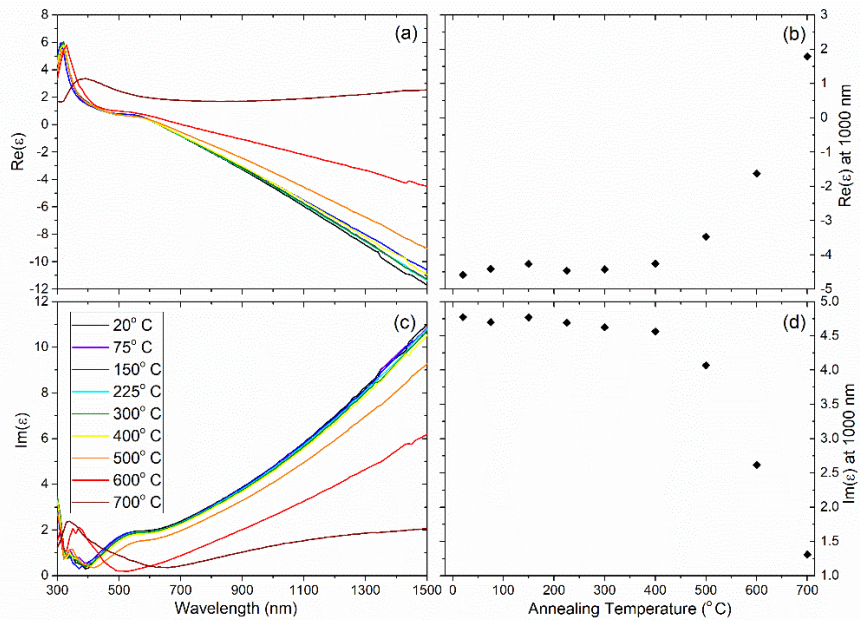

Fig. 9. Real (a) and imaginary (c) parts of the dielectric permittivity for sample annealed in air at temperatures up to 700° C. (b) and (d) show respectively the values of the real and imaginary parts of the permittivity measured at 1000 nm after each annealing process.

Figure 9 shows again the measured changes in the optical properties of the sample. However, in this case significant changes are not observed in the dielectric permittivity before annealing at 500° C while metallic behaviour is not lost altogether until annealing at 700° C, marking a substantial improvement on the sample produced without annealing in nitrogen. It can therefore be understood that the presence of the additional crystalline phase in the SMO thin film acts to suppress the high-temperature degradation mechanism, namely the loss of crystallinity, present in single phase SMO samples.

This improvement in high temperature stability is of particular significance when considering the specific plasmonic applications to which SMO may be best suited. Lalisse et al. recently introduced two figures of merit for the comparison of plasmonic materials, namely, the Faraday and Joule factors defined by Equation 1 and 2 below [17].

$$Fa = \left|\frac{E_{in}}{E_0}\right|^2 = 9\left|\frac{\varepsilon_s}{\varepsilon + 2\varepsilon_s}\right|^2 \quad (1)$$

$$Jo = \frac{e\varepsilon''}{n_s}\left|\frac{E_{in}}{E_0}\right|^2 = \frac{9e\varepsilon''}{n_s}\left|\frac{\varepsilon_s}{\varepsilon + 2\varepsilon_s}\right|^2 \quad (2)$$

The Faraday and Joule factors respectively quantify the maximum electric field enhancement of a plasmonic nanoparticle and its ability to locally generate heat as functions of the dielectric permittivity of a given material. In defining these figures of merit, the authors provided a comparison of a range of plasmonic materials, in which it was noted that TiN, the subject of much recent research [18-23], exhibits favourable broadband heat generation capabilities suitable for applications in thermoplasmonics, despite a relatively poor near-field enhancement. Figure 10 shows how the Faraday and Joule factors calculated for 100 nm samples of SMO (produced both with and without a one-hour anneal in $N_2$) compare with the values associated with Au and TiN.

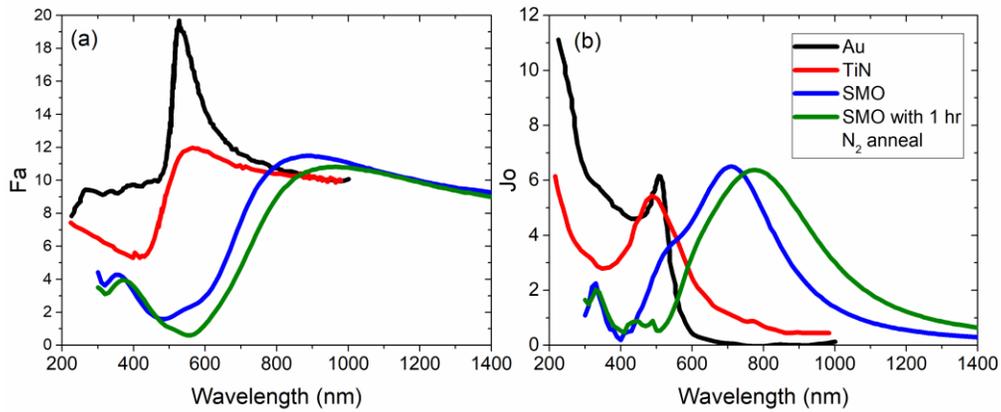

Fig. 10. Faraday (a) and Joule (b) factors for Au,[18] TiN,[18] and SMO, deposited both with and without a one-hour anneal in $N_2$.

It can be seen from Figure 10 that, as with the ellipsometry results of Figure 5, the nitrogen annealing process induces a slight red shift in the plasmonic resonance of SMO. However, in both cases, SMO, like TiN, is shown to be poorly suited to applications requiring a significant near-field enhancement. On the other hand, for applications in thermoplasmonics, SMO appears exceptionally well suited, offering heat generation over a broader bandwidth than TiN, and of a greater intensity on resonance. Additionally, one may note that the resonance observed here is centered in the biological transparency window at around 800 nm, suggesting the possibility of applications in photothermal therapies [24].

**4. Conclusion**

To conclude, strontium molybdate thin films were deposited by pulsed laser deposition onto STO substrates. Following film growth, samples were either cooled to room temperature under vacuum conditions or annealed in nitrogen prior to cooling. From XRD and spectroscopic ellipsometry measurements it has been shown that the nitrogen annealing process gives rise to an additional crystalline phase in the SMO thin film, leading to a reduction in optical losses. Subsequently, samples produced both with and without the nitrogen annealing process were annealed, ex-situ, in air to establish their maximum operating temperature. SMO films deposited without annealing in nitrogen were found to be highly unstable at elevated temperatures, with significant changes observed in the optical properties of samples when annealing at temperatures greater than 300° C, due to a loss of crystallinity in the film. However, this mechanism was found to be suppressed for samples which were first annealed in nitrogen, with metallic behaviour retained until annealing at temperatures above 600° C. Furthermore, through consideration of the Faraday and Joule figures of merit, SMO is shown to be well suited to thermoplasmonic applications, offering plasmonic heat generation over a broader bandwidth and of a greater intensity than TiN, previously identified as a promising material for such applications. The results therefore serve to identify the specific plasmonic applications to which SMO is best suited, along with showing considerable improvements in the material's chemical and thermal stability, and thus constitute a significant advancement towards SMO-based plasmonic devices.

**Funding**

Engineering and Physical Sciences Research Council (EPSRC) Reactive Plasmonics Programme (EP/M013812/1); Henry Royce Institute through EPSRC grant EP/R00661X/1.

**Acknowledgments**

The manuscript was written through contributions of all authors. P.K.P., M.P.W. and A.P.M. conceived and designed the research. M.P.W., S.F. and A.R., carried out the experiments. A.R. acknowledges the support from Imperial College London for her Imperial College Research Fellowship. S.A.M. further acknowledges the Lee-Lucas Chair.